\documentclass[conference]{IEEEtran}
\IEEEoverridecommandlockouts
\usepackage{cite}
\usepackage{amsmath,amssymb,amsfonts}
\usepackage{algorithmic}
\usepackage{graphicx}
\usepackage{textcomp}
\usepackage{xcolor}
\usepackage{url}
\usepackage{array,multirow}
\usepackage{multirow}
\usepackage{siunitx}
\usepackage{scalerel,stackengine}
\usepackage{makecell}
\usepackage{multirow}
\usepackage{tabularx}
\usepackage[caption=false, font=footnotesize]{subfig} 
\usepackage[inline]{enumitem}
\usepackage{numprint}

\usepackage{fixme}
\fxsetup{
    status = draft, 
    author = ,
    layout = inline, 
    theme = color}
\definecolor{fxnote}{rgb}{0.0000,0.6000,0.0000} 
\definecolor{fxwarning}{rgb}{1.0000,0.5490,0.0000} 
\definecolor{fxerror}{rgb}{1.0000,0.2706,0.0000}

\graphicspath{{./figures/PNG/}}

\DeclareGraphicsExtensions{.png,.jpg}

\usepackage{glossaries-extra}

\setabbreviationstyle[acronym]{long-short}
\setabbreviationstyle{short-long}
\glssetnoexpandfield{long}
\glssetnoexpandfield{longpl}
\newcommand*{\parentabbr}[2]{%
  \ifglsused{#1}{#2\glsentryshort{#1}}{\protect\glsunset{#1}#2\glsentrylong{#1}}%
}

\newacronym{mtc}{MTC}{Machine Type Communications}
\newacronym{mmtc}{mMTC}{\parentabbr{mtc}{massive }}
\newacronym{m2m}{M2M}{Machine to Machine}
\newacronym{iot}{IoT}{Internet of Things}
\newacronym{nbiot}{NB-IoT}{\parentabbr{iot}{Narrowband }}
\newacronym{gnss}{GNSS}{Global Navigation Satellite Systems}
\newacronym{esa}{ESA}{European Space Agency}
\newacronym{mcs}{MCS}{Modulation and Coding Scheme}
\newacronym{ue}{UE}{User Equipment}
\newacronym{ra}{RA}{Random Access}
\newacronym{rach}{RACH}{Random Access Channel}
\newacronym{nprach}{NPRACH}{Narrowband Physical Random Access Channel}
\newacronym{rv}{RV}{Random Variable}
\newacronym{pdf}{pdf}{probability density function}
\newacronym{gw}{GW}{Gateway}
\newacronym{geo}{GEO}{Geostationary Earth Orbit}
\newacronym{meo}{MEO}{Medium Earth Orbit}
\newacronym{leo}{LEO}{Low Earth Orbit}
\newacronym{fov}{FOV}{Field Of View}
\newacronym{fsl}{FSL}{Free Space Loss}
\newacronym{ul}{UL}{Uplink}
\newacronym{dl}{DL}{Downlink}
\newacronym{db}{dB}{decibel}
\newacronym{snr}{SNR}{Signal-to-Noise Ratio}
\newacronym{sir}{SIR}{Signal-to-Interference Ratio}
\newacronym{sinr}{SINR}{Signal-to-Interference plus Noise}
\newacronym{ecef}{ECEF}{Earth Centered Eart Fixed}
\newacronym{kpi}{KPI}{Key Performance Indicator}
\newacronym{phy}{PHY}{physical layer}
\newacronym{lut}{LUT}{Look Up Table}
\newacronym{awgn}{AWGN}{Additive White Gaussian Noise}
\newacronym{3gpp}{3GPP}{Third Generation Partnership Project}
\newacronym{nr}{NR}{New Radio}
\newacronym{ntn}{NTN}{Non-Terrestrial Network}
\newacronym{ntns}{NTNs}{Non-Terrestrial Networks}
\newacronym{ml}{ML}{Machine Learning}
\newacronym{uas}{UAS}{Unmanned Aircraft System}
\newacronym{isl}{ISL}{Inter-Satellite Links}
\newacronym{nn}{NN}{Neural Networks}

\def\BibTeX{{\rm B\kern-.05em{\sc i\kern-.025em b}\kern-.08em
    T\kern-.1667em\lower.7ex\hbox{E}\kern-.125emX}}
    
\begin{document}

\title{RAN Functional Splits in NTN: Architectures and Challenges}

\author{\IEEEauthorblockN{Riccardo Campana\IEEEauthorrefmark{1}, Carla Amatetti\IEEEauthorrefmark{1}, and
Alessandro Vanelli-Coralli\IEEEauthorrefmark{1}}
\IEEEauthorblockA{\IEEEauthorrefmark{1}Dept. of Electrical, Electronic, and Information Engineering (DEI), Univ. of Bologna, Bologna, Italy}}
\maketitle

\begin{abstract}
While 5G networks are already being deployed for commercial applications, Academia and industry are focusing their effort on the development and standardization of the next generations of mobile networks, \textit{i.e.}, 5G-Advance and 6G. Beyond 5G networks will revolutionize communications systems providing seamless connectivity, both in time and in space, to a unique ecosystem consisting of the convergence of the digital, physical, and human domains. In this scenario, Non-Terrestrial Networks (NTN) will play a crucial role by providing ubiquitous, secure, and resilient infrastructure fully integrated into the overall system. 
The additional network complexity introduced by the third dimension of the architecture requires the interoperability of different network elements, enabled by the disaggregation and virtualization of network components, their interconnection by standard interfaces and orchestration by data-driven network artificial intelligence. 
The disaggregation paradigm foresees the division of the radio access network in different virtualized block of functions, introducing the concept of functional split. Wisely selecting the RAN functional split is possible to better exploit the system resources, obtaining costs saving, and to increase the system performances. 
In this paper, we firstly provide a discussion of the current 6G NTN development in terms of architectural solutions and then, we thoroughly analyze the impact of the typical NTN channel impairments  on the available functional splits. Finally, the benefits of introducing the dynamic optimization of the functional split in NTN are analyzed, together with the foreseen challenges.

\end{abstract}

\glsresetall

\section{Introduction}\label{sec:intro}

\begin{figure}[t!]
    \centering
    \includegraphics[width=0.8 \columnwidth]{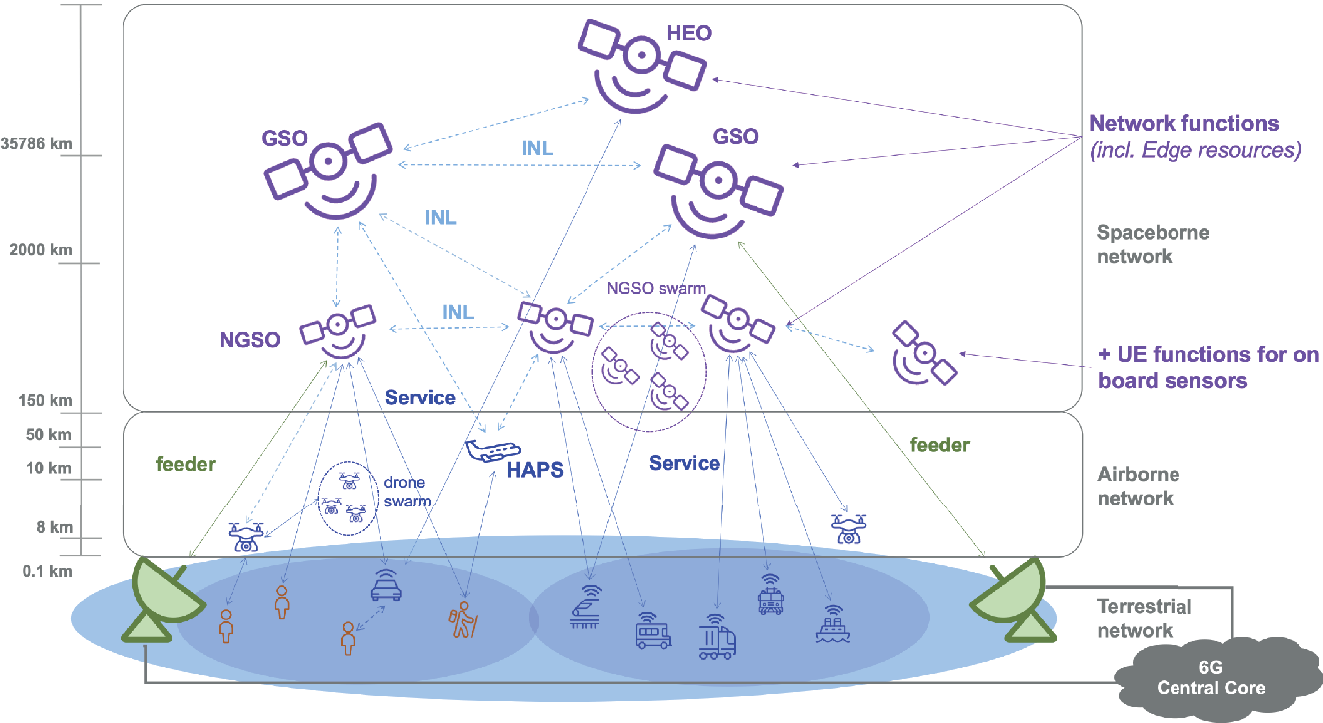}
    \caption{The foreseen multi-layer multi-dimensional architecture, \cite{guidotti2022path}.}
    \label{fig:3d_architecture}
\end{figure}

Non-Terrestrial Networks (NTN) have become a cornerstone in the evolution of 5G to 6G networks. 
Indeed, the Third Generation Partnership Program (3GPP) with its Rel.17 paved the way for the definition of a truly integrated Non-Terrestrial (NT) component into the terrestrial 5G system in order to complement 5G services in under-/un-served areas, improve the 5G service reliability and continuity for massive Machine Type Communications (mMTC), Internet of Things (IoT) devices, or for Mission Critical services, and improve the 5G network scalability by means of efficient multicast/broadcast resources for data delivery. Recently, 3GPP has started the definition of 5G-Advance (5G-A) NTN in Rel.18 through i) three Work Items (WIs) aimed at enhancing the features of NR and IoT NTN and NR via satellite backhaul; and ii) six Study Items (SIs) more related to the system aspects. With respect to Rel. 19, preliminary studies have been carried out in Technical Specification Group Services and System Aspects. In this context, it is clear that the interest and effort in the framework of 3GPP NTN will produce enhancements enabling increased performance and/or new capabilities for the NTN component in 5G-A and future 6G systems. Indeed, 6G is expected to create a fully connected world, where the physical domain is represented in high detail in the digital one, where it can be analyzed and also acted upon. The network would provide links between different domains through devices embedded everywhere as well as the infrastructure and the Artificial Intelligence (AI) of the digital world. In order to meet such demanding and challenging requirements, it is widely recognized that the unification of the terrestrial and non-terrestrial infrastructure components will be fundamental. Indeed, 6G systems foresee the joint design and optimization of Terrestrial Networks (TNs) and NTN in a unified and fully integrated multi-layered infrastructure.
The latter will combine terrestrial, airborne, such as High Altitude Platform Systems (HAPS) and Unmanned Aerial Vehicles (UAVs), and space-born network elements, \textit{i.e.,} satellites, networks elements for the envisaged convergence of the physical, human, and digital worlds. This multi-dimensional system, depicted in Figure \ref{fig:3d_architecture}, aims at unleashing the full potentiality of the 6G service performances (availability, resilience, coverage, throughput, and service rate) employing the NTN domain and considering its specific peculiarities and operational constraints.
Clearly, the third dimension of the architecture highly increases the complexity of the network, due to the required interoperability among the different nodes, while enabling a plethora of different services. Thus, multi-vendor interoperability, disaggregation, softwarization, and virtualization of Radio Access Network (RAN) functionalities  are fundamental enablers of such kind of architecture \cite{polese_understanding_2022}. 
This paradigm allows improved flexibility, hardware independence, and evolvability. In this framework, there are standardization fora (e.g., 3GPP, ITU, and ETSI) and companies alliances, such as O-RAN, TIP, and GSMA,  that strongly support these approaches. According to the standard, the disaggregation foresees the division of the RAN protocol stack into two sets of network functions, namely Central Unit (CU) and Distributed Unit (DU). The CU hosting the RRC and PDCP layers and the DU hosting the RLC, MAC, PHY, and RF layers, \cite{3gppTR38401}. This static disaggregation choice could be enhanced allowing more degrees of freedom in terms of: i) which parts of the stack are implemented in each unit according to the 8 possible splits, and ii) where each set of network functions is executed. Moreover, most network functions can be easily virtualized and moved to different nodes as long as latency and throughput requirements are met. The disaggregation allows the fine-tuning of the virtualized RAN bringing flexibility to RAN operations, potentially offering a cost-saving, and accommodating different use cases and applications in 5G-A systems. However, selecting the gNB functional split is challenging since each split has different delay requirements, initiates a different computing load to the CU and DUs, and induces a different data flow.
This motivated different works in the literature to assess the best functional split use cases and  the most efficient implementations \textit{i.e.,} \cite{FS_survey}, \cite{FFS_queuingModel}, and \cite{FFS_management}. 
However, the State of the Art addressing the 5G functional splits in NTN is limited. In our previous work \cite{oranntn}, we described the possible implementation of an NTN infrastructure based on the O-RAN approach.  In this paper, we firstly provide a discussion of the current 5G NTN development in terms of architectural solutions and then, we thoroughly analyze the impact of the typical NTN channel impairments  on the available functional splits to provide design criteria. Finally, the benefits of introducing the dynamic optimization of the functional split in NTN are analyzed, together with the introduced challenges.

\section{NTN System Architecture}\label{sec:sys_model}
\begin{figure}[t!]
    \centering
    \includegraphics[width=0.9 \columnwidth]{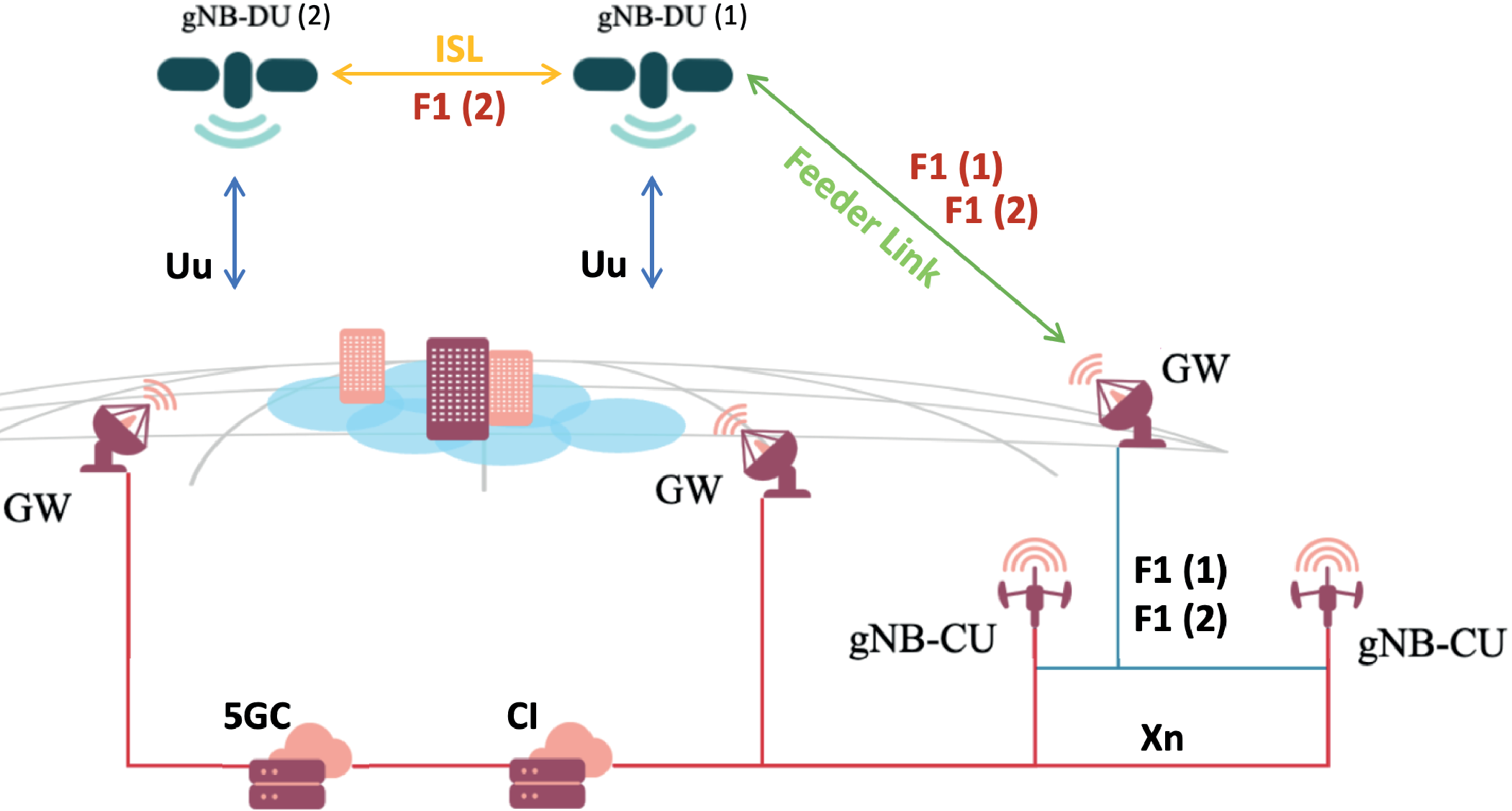}
    \caption{High-level system architecture.}
    \label{fig:syst_architecture}
\end{figure}
An NTN-based New Radio (NR) RAN architecture consists of NTN payloads, i.e., network elements on-board a satellite, NTN Gateways (GW), and a terrestrial segment.  . The GW interconnects the payload to the terrestrial segment via a feeder link, as shown in Figure \ref{fig:syst_architecture}. The terrestrial segment is constituted of the 5G Core network (5GC) and a Centralized Intelligence (CI). The role of CI is to gather information about the network status and to use it to select and implement the best network configurations that optimize network performances.\\
Different NTN elements can orbit the Earth at different altitudes providing services to the on-ground users. These elements represent the NTN access segment. If the payload is able to perform the radio frequency filtering, frequency conversion and amplification as well as demodulation/decoding, switching and/or routing, it is called regenerative. Performing such functions is equivalent to have all or part of the gNB protocol stack on the NTN platform. Contrary, if the payload implements only frequency conversion, filtering, and
amplification, then it only forwards the RF signals between the GW and UEs. In this case, the payload is called transparent and the gNB is conceptually located at the GW. \\
Depending on the antenna, the satellite is able to serve a portion of the on-ground coverage area by means of a single-beam or multi-beam system. The beams can be fixed or moving. In particular, on the one hand, the payload can be equipped so as to be able to steer the on-board antenna radiation pattern in order to always cover the same on-ground area, meaning that the beams are fixed. On the other hand, the area covered by a satellite will move accordingly with the satellite's movement in its orbit, i.e., the beams are moving with the satellite since the coverage center is always located at the sub-satellite point. This second case refers to the so-called moving beams. \\
Upcoming 5G-A NTN releases by 3GPP are likely to be based on a regenerative payload \cite{eager}. When the full protocol stack of the gNB is embarked on board, the Uu Air-Interface is present on the user link (i.e., the link connecting the users and the satellite). On the feeder link between the satellite and the GW, the interface carried is the Next Generation (NG) logical interface, which can be transported by any Satellite Radio Interface (SRI), such as DVB-S2, DVB-S2X, or DVB-RCS2 as long as specifically signaling operations are guaranteed.
As for the functional split, different operations will be performed on-ground (CU) and on-board (DU), as detailed below. According to 3GPP TS 38.401 \cite{3gppTR38401}, it shall be noticed that: i) a gNB can be split into a CU and one or more DUs; ii) each gNB-DU can be connected to a single gNB-CU, while a single gNB-CU is able of managing multiple gNB-DUs; and iii) the on-ground gNB-CU and the on-board gNB-DU are connected through the F1 interface.   
This interface is again logical and, thus, can be transported by means of any SRI ensuring specific signaling operations. The F1 interface requires a persistent connection between the gNB-DU and the gNB-CU. Even if 3GPP defines a strict set of RAN functions to be implemented in the CU and RU, in the scope of this work we assume that there is no limitation in the number and type of RAN functions that can be implemented in the CU and DU. This assumption enables the implementation of all 8 functional splits between the two units. Accordingly, we assume the F1 interface to be able to handle the signalling and Protocol Data Uniss (PDU) delivery of all the functional splits.
The physical link between a LEO satellite and the serving CU on-ground is guaranteed exploiting Inter-Satellite Links (ISL) in the case they are not in direct visibility.
It shall be considered that each gNB is capable of managing a few tens of beams. However, in the case of multi-beam NTN systems, each satellite may generate hundreds of beams. Therefore, in order to manage the NTN node, multiple gNBs (or part of it) might be needed on board. \\
Finally, in the user segment, we only consider users directly connected to the NTN elements.

\begin{table*}[t]
  \centering
  \caption{Overview of delay and bandwidth requirements, pros and cons of functional splits, informative data from \cite{3gppTR38473}.}
\begin{tabular}{l l l l l}
\hline 
Split & Transport latency & DL/UL Data rate & Pros & Cons \\
\hline 
1: $\mathrm{RRC} / \mathrm{PDCP}$ & 10 $ms$ & $\sim 4 / 3$ Gbps & $\begin{array}{l}\text {User data close to tx point.} \\ \text {Benefits for edge computing.}\end{array}$ & $\begin{array}{l}\text {Low shared processing gain.} \\ \text {Few DUs connected to CU.}\end{array}$\\

2: $\mathrm{PDCP} / \mathrm{RLC}$ & 1.5-10 $ms$ & $4 / 3$ Gbps & $\begin{array}{l}\text {User plane separation and traffic} \\ \text {aggregation.}\end{array}$ & $\begin{array}{l}\text {Security configuration issues.} \end{array}$ \\

3: Intra-RLC & 1.5-10 $ms$ & $\sim 4 / 3$ Gbps & $\begin{array}{l}\text {Traffic aggregation and better flow} \\ \text {control. Potential handling of more} \\ \text {connected mode UEs.}\end{array}$ & $\begin{array}{l}\text {Latency requirements and dupli-} \\ \text {cation of buffers.}\end{array}$ \\

4: RLC/MAC & $\sim$ 100 $\mu s$  & $\sim 4 / 3$ Gbps & $\begin{array}{l}\text {Low and cell-load dependant data rate.} \end{array}$ & $\begin{array}{l}\text {Separation of RLC and MAC.} \end{array}$ \\

5: Intra-MAC & hundreds of $\mu s$ & $\sim 4 / 3$ Gbps & $\begin{array}{l}\text {Traffic aggregation and better in-} \\ \text {terference management.}\end{array}$ & $\begin{array}{l}\text {Additional scheduling complexity.} \\ \end{array}$ \\ 

6: MAC/PHY & 250 $\mu s$ & $\sim 4 / 5$ Gbps & $\begin{array}{l}\text {Traffic aggregation,} \\ \text {Centralized scheduling.} \\ \end{array}$ & $\begin{array}{l}\text {Stringent timing between} \mathrm{CU} \text {and} \\ \text {DU.}\end{array}$ \\

7: Intra-PHY & 250 $\mu s$ & $\sim 22.2 / 86$ Gbps & $\begin{array}{l}\text {Implementation of advanced receivers.} \\ \end{array}$ & $\begin{array}{l}\text {In-band protovcol for PRB allocation.} \\ \end{array}$ \\

8: PHY/RF & 250 $\mu s$ & 157.3/157.3 Gbps & $\begin{array}{l}\text {More efficient resource manage-} \\ \text {ment. Improvement of RF/PHY} \\ \text {scalability.}\end{array}$ & $\begin{array}{l}\text {High requirement in fronthaul for} \\ \text {latency and bandwidth.}\end{array}$ \\
\hline
\end{tabular}
\label{table:1}
\end{table*}

\section{Functional Split Options}
Several functional split options are available and each one is optimum for specific use cases. Currently, those considered as the most relevant are option 2 (RRC/PDCP split), Option 6 (MAC/PHY split), and Option 7 (Low PHY/High PHY split). However, it is worthwhile highlighting that only Option 2 is supported by 3GPP; the other options have not been standardized by 3GPP but they may be implemented. The main functional split options are listed in Table \ref{table:1} together with the pros and cons of implementing a specific functional split in NTN, the maximum one-way latency, and the required F1 interface bandwidth. The options from 1 to 8 are also described in the following subsections for completeness, \cite{FS_survey}.

\subsection{Option 8: RF/PHY}
Option 8 splits the gNB between the Physical (PHY) layer and the Radio Frequency (RF) section. Indeed, only the RF sampler and the upconverter are left in the DU, resulting in a very simple DU, while all the remaining functions are centralized in the CU assuring the highest level of centralization. With this configuration, time In-phase - Quadrature (IQ samples are encapsulated in a protocol and delivered through the F1 interface connecting the CU and DU. Thus, the data rate on the F1 is constant, very high, and scales with the number of DU and antennas. 

\subsection{Option 7: Intra PHY}
Option 7 splits the gNB at the PHY layer and is divided into options 7.1, 7.2, and 7.3 depending on the specific PHY functions that are centralized.

\begin{itemize}
    \item Option 7.1 (Low PHY): The Fast Fourier Transformation (FFT) is decentralized in the DU. This implies a reduction in the required interface data rate compared to option 8, even if it is still constant since the resource element mapping is performed in the CU. 
    \item Option 7.2 (Low PHY/High PHY): The precoding and resource element mapper and decentralized in the DU. This split option reduces the required data rate since the F1 interface transports subframe symbols. Furthermore, the data rate becomes dependent on the required user traffic, implying a sensible data rate reduction in low traffic load contexts.
    \item Option 7.3 (High PHY): The scrambling, modulation, and layer mapping are decentralized in the DU. The signal delivered through the interface is modulated, sensibly reducing the required data rate proportionally to the modulation scheme used.
\end{itemize}
\subsection{Option 6: MAC/PHY}
Option 6 splits the gNB between the Medium Access Control (MAC) and PHY layers, resulting in all the physical processing handled in the DU and the MAC scheduler centralized in the CU. The F1 interface delivers transport blocks and this leads to a considerable reduction in the required data rate compared to higher split options. Indeed, the load on the interface is proportional to the cell load.

\subsection{Option 5: Intra MAC}
Option 5 splits the gNB inside the MAC layer, centralizing the overall scheduling task in the CU and leaving time-critical processing in the DU. The Hybrid Automatic Repeat Request (HARQ) procedures and the functions where performances are proportional to the latency are instantiated in the DU, meaning a drastic reduction of the interface latency constraints. On the other hand, many of the computationally critical tasks are left in the DU, reducing the benefits of shared processing.

\subsection{Option 4: RLC/MAC}
Option 4 splits the gNB between the Radio Link Control (RLC) and MAC layers. In this configuration, the scheduler is decentralized in the DU and it is distant from the closely related RLC functions leading to performance degradation. The data rate is lower and cell load dependent.

\subsection{Option 3: intra RLC}
Option 3 splits the gNB between high and low RLC layer, instantiating the RLC segmentation functions in the DU and the Automatic Repeat Request (ARQ) in the CU. This configuration facilitates the interconnection of multiple MAC entities with a common RLC entity. It also reduces the latency constraints since the scheduling computation is decentralized in the DU.

\subsection{Option 2: RLC/PDCP}
Option 2 foresees the Packet Data Convergence Protocol (PDCP) and RRC functions centralized in the CU while all the other gNB functions are performed in the DU. In this configuration all real-time functions are located in the DU, meaning very relaxed F1 interface latency requirements. Further, the centralized functions enable the dual connectivity feature. On the other hand, the vast majority of computational complexity is kept in the DU, implying a marginal computational multiplexing gain in the CU.

\subsection{Option 1: PDCP/RRC}
Option 1 implements all user plane functions in the DU, leaving only the Radio Resource Control (RRC) function inside the CU. This split offers marginal performance gains compared to the monolithic gNB implementation and imposes a limitation on the number of DUs connected to the same CU.

\section{NTN-imposed F1 Interface Challenges}

The interface connecting CU and DU must comply with strict requirements in terms of data rate, maximum delay, and reliability, as shown in Table ~\ref{table:1}. 
As extensively discussed both in the literature and in the standardization fora, satellite links are characterized by longer delays and higher losses than terrestrial ones. Thus, in this Section, we aim at describing the effects of these impairments on the realization of the functional splits. 

\subsection{Interface Capacity}

\begin{table}[t!]
\centering
\caption{Link budget and capacity analyses}
\label{tab:lb_analysis}
\resizebox{\columnwidth}{!}{%
\begin{tabular}{|c|cccc|}
\hline
\multicolumn{1}{|l|}{\multirow{2}{*}{}} & \multicolumn{1}{c|}{W}        & \multicolumn{1}{c|}{Q/V}    & \multicolumn{1}{c|}{Ku}        & Ka     \\ \cline{2-5} 
\multicolumn{1}{|l|}{}                  & \multicolumn{1}{c|}{83.5 GHz} & \multicolumn{1}{c|}{47 GHz} & \multicolumn{1}{c|}{28.75 GHz} & 17 GHz \\ \hline
Target ModCod                            & \multicolumn{4}{c|}{256 APSK 3/4}                                                                     \\ \hline
Spectral efficiency                     & \multicolumn{4}{c|}{5.9 {[}bit/s//Hz{]}}                                                              \\ \hline
Target C/N                              & \multicolumn{4}{c|}{24.02 {[}dB{]}}                                                                    \\ \hline
Bandwidth                               & \multicolumn{2}{c|}{22 GHz} & \multicolumn{2}{c|}{1 GHz}          \\                                      \hline                            
Target C/N0                             & \multicolumn{4}{c|}{112.4 {[}dBHz{]}}                                                                 \\ \hline
Required Ptx & \multicolumn{1}{l|}{59.8 {[}dBW{]}} & \multicolumn{1}{l|}{34.8 {[}dBW{]}} & \multicolumn{1}{l|}{6.07 {[}dBW{]}} & \multicolumn{1}{l|}{0.17 {[}dBW{]}} \\ \hline
Capacity                                & \multicolumn{2}{c|}{156.4 {[}Gbit/s{]}} &\multicolumn{2}{c|}{7.81 {[}Gbit/s{]}}  \\ 
\hline
\end{tabular}}
\end{table}

The capacity supported by the F1 interface is a fundamental aspect to analyze in the design of a functionally split RAN. In terrestrial networks, this interface is usually implemented through  high-capacity Ethernet networks or dedicated fiber optic links, guaranteeing adequate performance and reliability. In order to introduce the functional split concept in NTN networks, the F1 interface must be implemented through the NTN feeder link. That type of link is much less reliable and usually has less capacity than the terrestrial network links for which the functional split has been initially designed.
As investigated in the previous section, the required capacity changes notably in the different split options. The third column in Table \ref{table:1} reports the maximum capacity needed per split option assuming a gNB 100 MHz channel bandwidth, 256 QAM modulation with the highest spectral efficiency, corresponding to index 27 of the available Modulation and Coding Scheme (ModCod), and 8 MIMO layers. Options 1 to 5 require a maximum data rate of 4 Gbps in DL and 3 Gbps in UL, with option 6 requiring an increased UL data rate of 5 Gbps. In these cases, the maximum data rate is reported even if the actual data rate needed is a function of the traffic required by the users. Increasing the split option to 7 and 8, the requirement increases drastically reaching 157.3 Gbps in UL and DL for option 8. Thus, the available data rate on the feeder link becomes a critical parameter in selecting the optimal functional split option of the specific NTN system. Table ~\ref{tab:lb_analysis} shows the feeder link bandwidth and the transmission power necessary to meet the requirements of the different splits in terms of data rate, where 
the DVB-S2X is used as a radio air interface for the feeder-link and the highest spectral efficiency ModCods are selected. Since the air interface on the F1 has not been defined for NTN, we choose the DVB-S2X as it is optimized specifically for satellite data transmissions. 
 With the aim to define the required transmission power to reach the spectral efficiency, a link budget computation has been performed considering the losses defined in \cite{guidotti2022feeder} for the considered frequency bands. As shown in Table ~\ref{tab:lb_analysis}, a very high transmission power is required to counteract the losses in the high frequencies, providing large bandwidths. 
Splits 1-6 can be satisfied in lower frequencies bands, lowering the transmission power. 
It is worth highlighting that requirements in Table ~\ref{table:1} are valid for one gNB. As per 3GPP specification \cite{3gppTS38_213}, a maximum number of 64 beams for 5G NR is available. However, a satellite might  generate an higher number of beams. Thus, if a one to one mapping between the 5G NR beams and the satellite beams is followed, then several DUs should be instantiated onboard the NTN node. Therefore, assuming the maximum traffic load on each DU, the aggregate traffic delivered through the feeder link increases linearly with the number of DUs. Consequently, given an available feeder link data rate, a trade-off is imposed between the functional split option and the number of DUs that can be instantiated onboard.

\begin{figure}[t]
  \centering
  \begin{minipage}[b]{0.45\textwidth}
  \subfloat[\label{subfig:user_link}]{{\includegraphics[width= 1\linewidth]{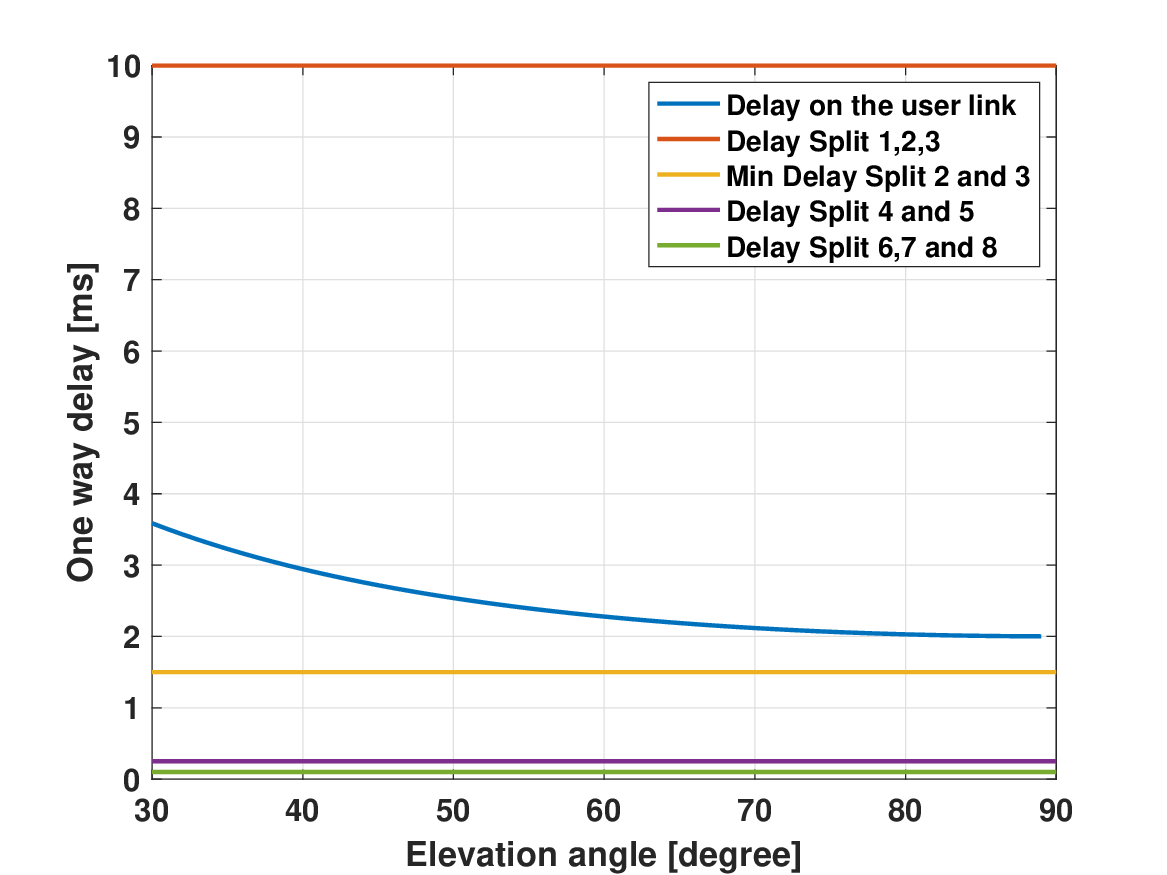}}}
    \hfill
    \subfloat[\label{subfig:sfeeder_link}]{
    \includegraphics[width=1\linewidth]{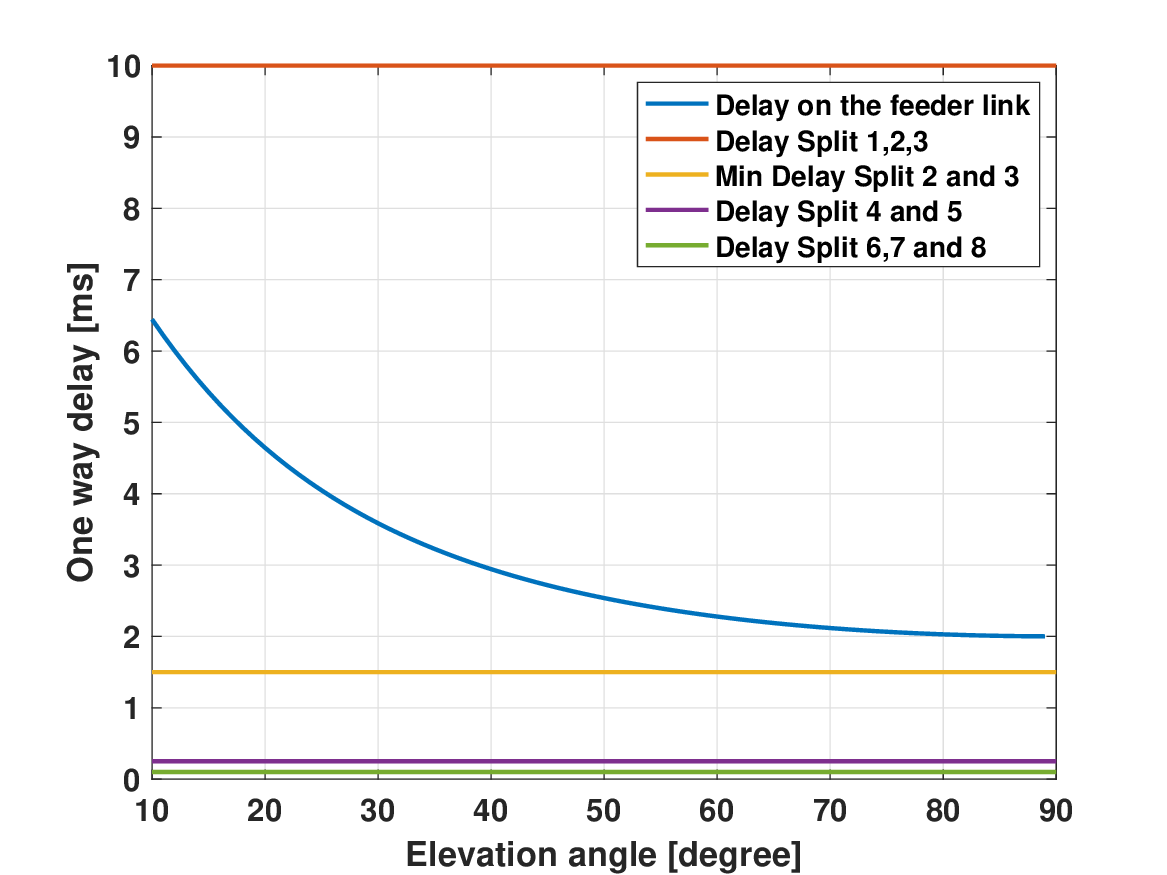}}
    \end{minipage}
  \caption{Maximum one-way delay over the user link varying the elevation angle of the beam center vs the delay of each split (a) and one-way on the feeder link (b). LEO at 600 km of altitude and 50 km of beam diameter.}
  \label{fig:delay}
\end{figure}

\subsection{Interface Latency}

RAN functional blocks require tight collaboration among them in order to deliver a good quality of service. Indeed, the communication latency limit between some RAN functions is very stringent, due to functional timers. The total latency comprises the internal computational time of the RAN functions and the PDU transport latency. Deploying the functions at different points of the network means increasing the PDU transport latency component of the delay, making the distance between the network nodes a parameter to be strictly controlled.
In the reference architecture, the RAN functions are split between the on-ground component and the onboard one. Thus, the F1 interface must be implemented through the SRI on the feeder link. In this case, the PDU transport latency coincides with the channel delay. 
The maximum channel delays in each functional split option, as reported for information by 3GPP in TR 38.801 \cite{3gppTR38473}, are listed in the second column of table \ref{table:1}. Options 1 to 3 have relaxed latency limits since they are upper-bounded only by the maximum RAN-5GC interface delay. In higher functional splits the latency limit is more stringent. Option 4 is limited by the scheduling interval between RLC and MAC. Considering the time for MAC scheduling and for RLC PDU segment concatenation, the maximum latency of option 4 is approximately 100 $\mu s$. In options 5 to 8 the most challenging timer to be met refers to the HARQ process. Indeed, it must be completed in maximum $5 ms$. Taking into account the processing delays and the transmission time to the UE, the time left for the F1 transport delay is approximately 250 $\mu s$.
Figure \ref{fig:delay} shows the maximum channel delay on the user link (\ref{subfig:user_link}) varying the elevation angle of the beam center, while Fig. ~\ref{subfig:sfeeder_link} provides the one-way channel delay on the feeder link. The minimum elevation angles considered in the user and feeder links are respectively $30^{\circ}$ and $10^{\circ}$, as specified in \cite{3gppTR38821}.  The graph refers to a scenario with LEO satellites orbiting at 600 Km of altitude and generating beams of 50 km diameter.
Comparing the channel delays of our reference scenario with the maximum F1 interface delays, it is clear that NTNs impose a real challenge to the implementation of the functional split paradigm. Indeed, options 1 to 4 can support a delay compliant with LEO NTN systems, while higher-level splits impose a latency constraint that can not be met in current systems. In order to enable an efficient implementation of high-level splits, next-generation systems should: 
\begin{enumerate*}[label=\emph{\roman*})]
    \item drastically decrease the required computational time to leave more time for the PDU transport latency, or
    \item relax the latency constraint on the HARQ function.
\end{enumerate*}
As suggested by 3GPP in \cite{3gppTR38821}, the latency constraint on the HARQ function can be relaxed extending the maximum number of parallel HARQ processes from 16 to 32. To further relax the timing requirement, the already existing HARQ processes could be reused. In this case, the same HARQ process is connected to multiple data transmissions. Thus, the HARQ feedback becomes unnecessary and can be disabled. 

\subsection{FH Interface PDUs reliable delivery}
Focusing on LEO satellites, they orbit around the earth between 11 and 16 times per day, causing a constant hand-over in the GW serving the satellite DU through the feeder link.
However, 3GPP defines the F1 interface as a constantly available logical link between the CU and the DUs, \cite{3gppTR38473}. Since the F1 interface is delivered through the feeder link, even if we assume the network is dimensioned with a sufficient number of GWs distributed on the earth's surface and Inter Node Links (INLs), such that there is always a viable route between the CU and the DUs, there will still be the need for feeder link and INL handovers. The frequent changes in the GW serving the satellite may cause the late delivery or even the loss of a consistent number of PDUs.
To enable the reliable delivery of the F1 PDUs from the CU to the DUs on the satellite, the CI shall enable the management of the feeder link and INLs handovers directly at the network layer, providing reliable service to the application layer on which the F1 interface relays. Indeed, an Artificial Intelligence (AI) application implemented in the CI is able to foresee the availability of each physical link under its control and can centrally orchestrate the network elements in order to always route the F1 packets toward an active network path. In this way, no F1 packet is lost or delayed trying to deliver it on a link that is going to drop.
The main challenge to be tackled while implementing this solution is to serve the DUs from different CIs while orbiting around the earth, since the single CI has not a global view of the network. This is actually a challenging architectural design to address since it implies:  
\begin{enumerate*}[label=\emph{\roman*})]
    \item to implement over-dimensioned CIs to keep in memory the data and applications of all the RAN elements, even the ones not in visibility, or
    \item to exchange the data and applications between the CIs when it is needed, causing an increased load on the core network in charge of managing the CIs.
\end{enumerate*}

\section{Future trend: AI-enabled Functional Split Optimization}

In the previous section, we have thoroughly described the main challenges that need to be addressed to complete the first step of the realization of the function split into the NTN component. However, in order to enable the full integration of TN-NTN components, a further level of optimization is needed, which combines resource optimization and Artificial Intelligence in the RAN. \\
In a static system, the optimal functional split can be already determined in the system design phase. On the contrary, the NTN architecture changes morphology in a highly dynamic way, thus, a dynamic optimization of its functional split could be beneficial.
In this respect, the NTN architecture shall enable a system-aware and proactive functional split optimization. Indeed, the CI will be in charge of computing the optimal functional split based on status data collected from the network and redeploying the network functions in the CU and DU according to it. 
One of the main optimization objectives to be investigated is the minimization of the on-board payload energy consumption.
Indeed, the satellite has limited available power, this implies that every single watt from batteries and the photovoltaic unit shall be wisely exploited. Moreover, not only the communication payload power is a scarce resource, but also it is not constant in time since it is a function of the satellite's position in the orbit and it depends on the instantaneous power required by the other satellite subsystems. In this framework, an AI application deployed in the CI will be able to select and implement the optimal functional split while guaranteeing an appropriate Quality of Service (QoS). Precisely, shifting RAN functions from the on-board DU to the on-ground CU frees up resources on the payload but increases the required feeder link performances and the latency of the CU-implemented RAN functions. 
This application operates by collecting data from the network about: 
\begin{enumerate*}[label=\emph{\roman*})]
    \item type and volume of requested user traffic;
    \item payloads computational power capabilities;
    \item Payloads instantaneous available power, and
    \item the  feeder link instantaneous data rate and transport latency.
\end{enumerate*}

Additionally, in system configurations with lower restrictions on power consumption, an additional optimization objective is the maximization of the exploitation of the feeder link, following the time-varying behavior of its performances. In this regard, the CIs will exploit the same KPIs of the previous case to proactively select the best-fitting functional split.\
The most challenging aspect of the dynamic functional split implementation is the high functional flexibility required on the payload. Indeed, that grade of flexibility that can be met by:
\begin{enumerate*}[label=\emph{\roman*})]
    \item relying on general-purpose computing processors, or
    \item implementing the single RAN functions on specialized and isolated hardware that can be individually activated.
\end{enumerate*}
The general-purpose computing technology currently available requires an amount of power not compliant with the common NGSO payload implementations, \cite{electronics11132024}. Additionally, the latter case implies high complexity design and poorly exploited payload hardware. This implies the current applicability of the dynamic functional split is limited to low-capacity services, but relying on the future implementation of low-power high-performance processors its applicability can be extended to every type of service.

\section{Conclusion}
The disaggregation, softwarization, and virtualization of Radio Access Network functions are the key enablers of the 6G networks. Since the latter foresees the unification of the terrestrial and non-terrestrial networks, it is of paramount importance to address the peculiar challenges of the satellite links, to efficiently design a flexible and optimized RAN. \\  
In this article, we have 
introduced the concept of functional split and investigated its exploitation in NTNs. We started with the description of the different functional split options. Consequently, we have identified the challenges imposed by the NTN system to the implementation of the F1 interface. For the F1 data rate, we assessed that lower-level functional splits can be delivered through the simulated SRI of 1GHz in Ku and Ka bands, while for splits 6 and 7, W and Q/V frequency bands are necessary to gain higher bandwidths. For the stringent F1 delay constraints, we have shown that only lower-level functional splits can be instantiated through a NTN feeder link without performing adaptations to the NR protocol. Precisely, we suggested that to enable the exploitation of higher-level functional splits it is fundamental to relax the timing requirements on the HARQ procedure, feature supported since 3GPP Release 17. Further, we underlined the importance of having a reliable delivery of CD-DU interface PDUs through the SRI. Finally, we presented the benefits of implementing a dynamic functional split in the context of NTN networks and we proposed an architectural solution to implement this dynamic change of spit option. 

\section{Acknowledgment}
This work has been funded by the 6G-NTN project, which received funding from the Smart Networks and Services Joint Undertaking (SNS JU) under the European Union’s Horizon Europe research and innovation programme under Grant Agreement No 101096479. The views expressed are those of the authors and do not necessarily represent the project. The Commission is not liable for any use that may be made of any of the information contained therein.
\bibliographystyle{IEEEtran}
\bibliography{IEEEabrv,biblio_RAML}

\end{document}